%% file: main.tex
\documentclass[sigplan,10pt]{acmart}

\setcopyright{none}
\renewcommand\footnotetextcopyrightpermission[1]{}

\input{helpers/packages}

\AtBeginDocument{%
  }

\setcopyright{acmlicensed}
\copyrightyear{2026}
\acmYear{2026}
\acmDOI{XXXXXXX.XXXXXXX}
\acmConference[ATC '26]{2026 ACM SIGOPS Annual Technical Conference}{November 16--18,
  2026}{Shatin, Hong Kong}
\acmISBN{978-1-4503-XXXX-X/2026/11}

\begin{document}

\title{\tool: Safely Extending the eBPF Compilation Pipeline with Native Operations}

\author{Yusheng Zheng$^{1*}$, Zhengjie Ji$^{2*}$, Weichen Tao$^{3}$, Hao Sun$^{4}$, Wei Zhang$^{5}$, Dan Williams$^{2}$, Andi Quinn$^{1}$}
\thanks{$^*$Equal contribution.}
\affiliation{%
  \institution{$^{1}$UC Santa Cruz \quad $^{2}$Virginia Tech \quad $^{3}$Telecom Paris \quad $^{4}$ETH Zurich \quad $^{5}$University of Connecticut}
  \country{}}
\renewcommand{\authorsaddresses}{}

\renewcommand{\shortauthors}{Zheng et al.}

\input{helpers/commands}
\input{sections/0-abstract}




\maketitle
\fancyhf{}
\fancyfoot[C]{\thepage}
\renewcommand{\headrulewidth}{0pt}

\input{sections/1-introduction}

\input{sections/2-background}
\input{sections/3-characterization}
\input{sections/4-bpfext}
\input{sections/5-kinsn}
\input{sections/6-implementation}
\input{sections/7-evaluation}

\input{sections/8-related-work}
\input{sections/9-conclusion}

\bibliographystyle{ACM-Reference-Format}
\bibliography{reference}

\end{document}

%% file: helpers/packages.tex
\usepackage{xspace}
\usepackage{listings}
\usepackage{xcolor}
\usepackage{subcaption}

\lstdefinestyle{asm}{
  basicstyle=\ttfamily\scriptsize,
  columns=fullflexible,
  keepspaces=true,
  commentstyle=\color{gray},
  morecomment=[l]{;},
  morecomment=[l]{//},
}

\usepackage[most]{tcolorbox}

%% file: helpers/commands.tex
\newcommand{\zj} [1]{\textcolor{blue}{{}#1}}
\newcommand{\note} [1]{\textcolor{red}{#1}}
\newcommand{\todo} [1]{\textcolor{red}{#1}}
\newcommand{\revise} [1]{\textcolor{orange}{#1}}
\newcommand{\revised} [1]{\textcolor{green!50!gray}{#1}}

\newcommand{\tool}{\textsc{Kops}\xspace}
\newcommand{\lang}{\textsc{EInsn}\xspace}

\newcommand{\citehere}[1][?]{%
  \textcolor{red}{[#1]}%
}

\newtcolorbox[auto counter]{finding}{
  enhanced, breakable, sharp corners,
  colback=black!4, colframe=black!75, boxrule=0.5pt,
  left=2.2mm, right=2.2mm, top=1.4mm, bottom=1.4mm,
  before skip=2.4mm, after skip=2.4mm,
}
\newcommand{\finlabel}{\textbf{Finding~\thetcbcounter:}\itshape\ }

%% file: sections/0-abstract.tex




\begin{abstract}
eBPF safely extends OS kernels in domains such as networking, observability, and security.
The safety comes from an in-kernel compilation pipeline where a verifier checks every program, and a kernel just-in-time compiler (JIT) translates the verified bytecode to native code.
The kernel keeps the JIT simple to stay trustworthy, translating one bytecode instruction at a time in a single pass.
This single-pass design misses optimization opportunities, so eBPF runs up to twice as slow as natively compiled code in our characterization.
Adding optimizations to the kernel JIT directly requires upstream acceptance and a long release cycle, enlarges the trusted computing base (TCB), and grows the per-architecture kernel code.

To address this, we present \tool, an extension interface that lets userspace compilers and kernel modules introduce new operations without modifying the kernel core, while keeping a minimal trusted computing base (TCB).
Each operation has two forms, a \emph{proof sequence} of vanilla eBPF instructions that the existing verifier checks and a \emph{native emit} of machine instructions that the JIT compiles.
Because the verifier checks the proof sequence, the native emit is the only per-operation addition to the TCB.
Hardware idioms are the lowest-hanging fruit for this interface. With \tool, we build \lang, seven operations such as rotate and conditional select that CPUs execute as single instructions.
Lean 4 proofs show that each native emit computes the same result as its proof sequence.
On x86-64 and ARM64, \lang speeds up eBPF microbenchmarks by up to 24\% and production applications by up to 12\%.
The same interface also supports whole-program native replacement, reaching 2.358$\times$ at the cost of a larger TCB.
\end{abstract}

%% file: sections/1-introduction.tex
\section{Introduction}
\label{sec:introduction}

Extended Berkeley Packet Filter (eBPF) safely extends OS kernels in domains including networking~\cite{cilium,katran}, observability~\cite{bpftrace,bcc}, security~\cite{krsi_lwn,tracee}, and scheduling~\cite{sched_ext_lwn}.
eBPF's safety comes from a compilation pipeline that verifies every program before execution.
A userspace compiler lowers each program to eBPF bytecode, which the in-kernel verifier checks for safety properties such as memory safety and termination.
The kernel's just-in-time compiler (the kernel JIT) then translates the verified bytecode to native code.
The kernel keeps the JIT simple to stay trustworthy, translating one bytecode instruction at a time in a single pass.

This single-pass design, however, is where eBPF loses performance.
Because the JIT processes each instruction in isolation, it cannot recognize multi-instruction patterns or apply cross-instruction optimizations.
The same C code runs up to twice as slow through the pipeline as when compiled directly to native code (\S\ref{sec:characterization}).
An optimizing compiler fed the same verified bytecode recovers most of the gap, confirming that the loss comes from the kernel JIT's per-opcode code generation.

Existing optimization work stays in userspace because adding optimizations to the kernel JIT is hard.
First, any change requires upstream acceptance and a long release cycle.
Second, enlarging the JIT enlarges the trusted computing base (TCB), because the verifier checks only the bytecode and cannot verify the native code the JIT emits.
Third, native code is architecture-specific, so a larger JIT grows the per-architecture kernel code.
The verifier and the JIT carry years of analysis and formal verification~\cite{nelson2020specification,vishwanathan2023verifying}, so any kernel change must stay small.
Existing optimization work therefore stays in userspace: dedicated eBPF optimizers such as Merlin~\cite{mao2024merlin}, K2~\cite{xu2021synthesizing}, and EPSO~\cite{zhu2025epso} rewrite programs within the eBPF instruction set, but cannot emit native instructions the instruction set lacks.

We present \tool, an extension interface that lets userspace compilers and kernel modules introduce new operations without modifying the kernel core.
Our key idea is to give each operation two forms, a \emph{proof sequence} of vanilla eBPF instructions that the existing verifier checks, and a \emph{native emit} of machine instructions that the JIT compiles.
A compile-time recognizer rewrites programs to use the new operations.
The verifier re-checks the proof sequence on every load.
Because the verifier checks the proof sequence, the native emit is the only per-operation addition to the TCB.
Neither the recognizer nor the proof sequence is part of the TCB, so a recognizer bug cannot make a verified program unsafe.

Hardware idioms are the lowest-hanging fruit for this interface.
These are operations such as rotate, conditional select, and bit extract that CPUs execute as single native instructions but that the eBPF instruction set expresses only as multi-instruction sequences.
The idioms that the kernel JIT compiles poorly form a small, bounded set, so a handful of operations suffices.
A 64-bit rotate with a variable shift, a single \texttt{ROL} instruction on x86-64, arrives at the kernel JIT as eight bytecode instructions and is emitted as 15 machine instructions.
With \tool, we build \lang, a set of seven such operations.
We prove in Lean 4 that, for each operation, the native emit has the same eBPF-visible effect as the proof sequence, carrying the verifier's guarantee to the native code.

We implement \tool in the Linux kernel and in userspace tooling, as a small change to the kernel core, a compile-time recognizer, and kernel modules that supply the operations.
\lang speeds up eBPF microbenchmarks by up to 24\% on x86-64 and 22\% on ARM64 over the unmodified pipeline, recovering up to 42\% of the gap to native code, and improves production application throughput by up to 12\%.
Native code size shrinks by 12--23\% with no change to the existing verifier analysis (\S\ref{sec:eval}).
The same \tool mechanism also supports whole-program native replacement, reaching 2.358$\times$ on Cilium at the cost of trusting the application's native code (\S\ref{subsec:native_upper_bound}).
A new operation ships as a kernel module, and the existing verifier analysis stays untouched.

\noindent\textbf{Contributions.}
We make the following contributions:
\begin{itemize}
    \item We present \tool, a mechanism that extends the eBPF compilation pipeline with new operations while reusing the existing verifier for safety (\S\ref{sec:mechanism}).
    \item With \tool, we build \lang, a set of seven hardware-idiom operations that the simple kernel JIT emits directly (\S\ref{sec:lang}).
    \item We prove in Lean 4 that each \lang native emit has the same eBPF-visible effect as the proof sequence the verifier checks, so the verifier's guarantee carries over to the native code (\S\ref{sec:formal-verification}).
    \item An implementation in the Linux kernel and userspace tooling (\S\ref{sec:implementation}) recovers up to 42\% of the eBPF-native gap in our evaluation (\S\ref{sec:eval}).
\end{itemize}

%% file: sections/2-background.tex
\section{The eBPF Compilation Pipeline}
\label{sec:background}

This section describes the three stages of the eBPF compilation pipeline, with attention to how much each stage optimizes the program and to the instruction-set boundary that every stage respects.
Every eBPF program reaches the kernel through the pipeline, shown in Figure~\ref{fig:ebpf}.
A user-space compiler produces bytecode, which the kernel verifies and JIT-compiles.
The compiled program then attaches to a hook such as XDP or tc.

\input{figures/sec-2-ebpf-pipeline}

\noindent\textbf{Source compilation.}
The user-space compiler does most of the pipeline's optimization.
A language front end such as Clang lowers the source to LLVM IR, and LLVM's middle end runs its full optimization passes.
The eBPF backend then lowers the result to bytecode restricted to the operations the instruction set defines.
Optimization research also targets this stage, where compile-time optimizers such as Merlin~\cite{mao2024merlin}, K2~\cite{xu2021synthesizing}, and EPSO~\cite{zhu2025epso} rewrite programs for speed and size.
The optimizers all stay within the instruction set, a boundary the verifier enforces.
The instruction set has no single-instruction form for hardware idioms.
Every optimizer at this stage, including LLVM, therefore emits the idioms as multi-instruction sequences.

\noindent\textbf{Verification.}
The verifier checks safety, changes the program little, and enforces the instruction set.
The verifier proves memory safety, termination, and similar properties by static analysis~\cite{gershuni2019simple}.
The analysis proceeds opcode by opcode, over the 123 opcodes the instruction set defines~\cite{rfc9669}.
An opcode the instruction set does not define is rejected.
The rejection makes the instruction set a hard boundary for every upstream optimizer, so the idioms reach the JIT as multi-instruction sequences.
After a program passes, the verifier applies only a few targeted rewrites, such as dead-code elimination.
A few systems also optimize at this load-time stage, such as KFuse~\cite{kfuse}, which merges verified programs within the instruction set.
The verifier then passes the program to the JIT.
The kernel also exposes \emph{kfuncs}, native kernel functions that eBPF programs may call, registered by loadable modules with BTF type metadata~\cite{kfuncs}; \tool reuses this registration infrastructure (\S\ref{subsec:kernel}).

\noindent\textbf{JIT compilation.}
The kernel's per-opcode JIT does almost no optimization.
The kernel maintains one such JIT for each supported architecture.
Translation proceeds one opcode at a time, with the eBPF registers \texttt{r0}--\texttt{r10} mapped to fixed native registers.
The only optimization choices are within one instruction's encoding, such as picking the shortest immediate form on x86-64.
The JIT runs no general cross-instruction optimization, such as global register allocation or instruction scheduling.
The simple design keeps the JIT small and trustworthy beside the verifier, a tradeoff that \S\ref{subsec:char-challenges} returns to.
The multi-instruction sequences the backend emits therefore survive into the machine code unoptimized, at a cost that \S\ref{sec:characterization} measures.

%% file: figures/sec-2-ebpf-pipeline.tex
\begin{figure}[t]
  \centering
  \includegraphics[width=\linewidth]{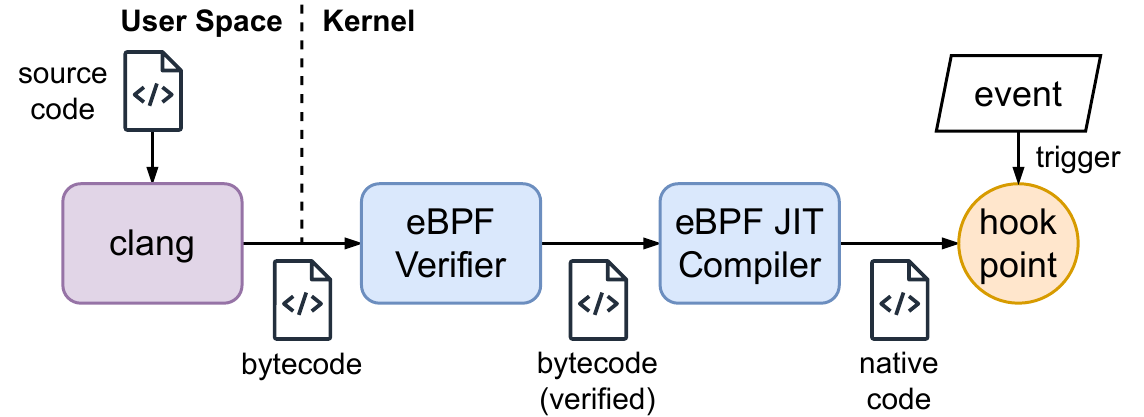}
  \caption{The eBPF execution pipeline.}
  \label{fig:ebpf}
\end{figure}

%% file: sections/3-characterization.tex
\section{Characterization of the eBPF-Native Performance Gap}
\label{sec:characterization}

The same source program runs 1.57$\times$ slower on x86-64, and 1.98$\times$ slower on ARM64, through the eBPF compilation pipeline than under direct native compilation (Table~\ref{tab:char-micro-summary}).
We characterize the gap on pure-bytecode microbenchmarks, computation kernels that use no helper calls or map accesses, through three questions:
\begin{itemize}
  \item \textbf{C-RQ1.} How much does in-kernel execution slow native code?
  \item \textbf{C-RQ2.} How much of the gap can an optimizing compiler recover from the same verified bytecode?
  \item \textbf{C-RQ3.} Which operations does the per-opcode kernel JIT compile poorly?
\end{itemize}

The three questions decompose the gap.
The first separates the cost of in-kernel execution from the quality of the machine code the pipeline generates.
Within code generation, the second weighs the per-opcode kernel JIT against the bytecode the JIT compiles.
The third identifies which operations the JIT compiles poorly, so a fix can target the operations.
The section closes with why the kernel cannot host an optimizing compiler (\S\ref{subsec:char-challenges}).

\subsection{Experimental Setup}
\label{subsec:char-method}

\noindent\textbf{Benchmarks.}
The corpus is 27 pure-bytecode microbenchmarks.
Each is a computation kernel from a real eBPF program in networking, tracing, or security.
The examples include a consistent-hash load-balancer lookup from \texttt{katran} and a SipHash-like 64-bit-rotation mixer.
The kernels use no helper calls or map accesses, so each measurement isolates instruction execution.
The gap therefore characterizes the bytecode-to-native code-generation quality; in production programs, helper and map time dilutes the instruction-execution share, so case study in \S\ref{sec:eval} measures the end-to-end effect separately.

\noindent\textbf{Execution paths.}
We compile each benchmark through four execution paths, the kernel eBPF JIT baseline plus the three paths in Table~\ref{tab:char-micro-summary}.
The paths vary the final backend and where the code runs.
The kernel eBPF JIT compiles the verified bytecode in the kernel and is the baseline.
In-kernel native and userspace native run native code from the same source, inside and outside the kernel.
Userspace LLVM-BPF recompiles the same bytecode with an optimizing compiler in userspace~\cite{zheng2025extending}.
Three pairwise comparisons attribute the gap.
In-kernel native against userspace native isolates in-kernel execution.
The kernel eBPF JIT against userspace LLVM-BPF isolates the per-opcode kernel JIT, given the in-kernel execution bound from C-RQ1.
Userspace LLVM-BPF against userspace native isolates the bytecode form.

\noindent\textbf{Metric.}
A path's speedup on a benchmark is the kernel eBPF JIT's median steady-state execution time divided by the path's time.
Each reported speedup is the geometric mean over the benchmarks.
Load, link, and compile time stay outside the speedup.

\input{tables/sec-3-micro-summary}

\subsection{C-RQ1: How Much Does In-Kernel Execution Slow Native Code?}
\label{subsec:gap}

In the kernel, native code runs about as fast as in userspace.
On x86-64, Figure~\ref{fig:char-pure-percase} shows in-kernel native at 1.55$\times$ and userspace native at 1.57$\times$ over the kernel eBPF JIT, a difference of about 2\%.
On ARM64 the difference is about 5\%, 1.88$\times$ against 1.98$\times$.
The two paths run native code from the same source and differ in where the code runs.
The spread therefore bounds the cost of in-kernel execution at about 2--5\%.
The cost is small, and the gap comes mainly from code generation.

\begin{finding}
\finlabel The eBPF-native gap comes from code generation. In-kernel execution slows native code by only about 2\% on x86-64 and 5\% on ARM64.
\end{finding}

\input{figures/sec-3-pure-bytecode-percase}

\subsection{C-RQ2: How Much Can an Optimizing Compiler Recover?}
\label{subsec:char-jit}

Code from an optimizing compiler on the same bytecode runs almost as fast as native code.
Userspace LLVM-BPF compiles the same verified bytecode that the kernel runs, but with an optimizing backend in place of the per-opcode kernel JIT.
As Table~\ref{tab:char-micro-summary} shows, userspace LLVM-BPF reaches 1.53$\times$ on x86-64 and 1.92$\times$ on ARM64.
Against userspace native at 1.57$\times$ and 1.98$\times$, the optimizing backend recovers 93\% and 94\% of the gap.
Given the in-kernel execution bound from C-RQ1, the recovered share measures the performance the per-opcode kernel JIT loses.
The remaining share of the gap, 7\% on x86-64 and 6\% on ARM64, comes from the bytecode form, which fixes choices such as the register set and calling convention before any compiler runs.

\begin{finding}
\finlabel The per-opcode kernel JIT is the dominant source of the generated code's slowness. An optimizing compiler recovers 93--94\% of the gap from the same verified bytecode. The verified bytecode carries most of what a compiler needs to generate fast code.
\end{finding}

\subsection{C-RQ3: Which Operations Does the Kernel JIT Compile Poorly?}
\label{subsec:char-idiom}

The per-opcode kernel JIT compiles hardware idioms poorly.
A hardware idiom has no single-instruction form in the eBPF instruction set, so the eBPF backend emits the idiom as several bytecode instructions (\S\ref{sec:background}).
The kernel JIT then translates each bytecode instruction separately into one or more machine instructions.
Figure~\ref{fig:jit-dump} shows a 64-bit rotate with a variable shift under the per-opcode kernel JIT and under direct native compilation.
The rotate arrives at the kernel JIT as eight bytecode instructions and is emitted as 15 machine instructions.
Direct native compilation uses a single rotate instruction, \texttt{ROL} on x86-64 or \texttt{ROR} on ARM64.

\input{figures/sec-3-jit-dump}

A small, bounded set of hardware idioms expands the same way under the kernel JIT.
The rotate is one of a handful of idioms with no single-instruction form.
Table~\ref{tab:isa-gap} lists the idioms, with the instruction counts along each path.
Across five idiom-stress probes on x86-64, small programs that each repeat one idiom, the kernel JIT emits 1.25--2.7$\times$ as many instructions as userspace LLVM-BPF for the same programs.
Code size shows the cost across the whole corpus.
In-kernel native machine code is about half the size of the kernel JIT output, 0.54$\times$ on x86-64 and 0.49$\times$ on ARM64, an aggregate that includes losses beyond the idioms.

\begin{finding}
\finlabel A significant share of the per-opcode JIT's loss comes from a small, bounded set of hardware idioms, such as rotate, wide load, and conditional select. The remaining loss comes from cross-instruction optimizations, such as register allocation and instruction scheduling, that are outside the scope of a per-idiom fix.
\end{finding}

\input{tables/sec-2-isa-gap}

\subsection{Why Not an Optimizing Compiler in the Kernel?}
\label{subsec:char-challenges}

The obvious response to the simple kernel JIT is to replace the JIT with an optimizing compiler.
An optimizing compiler in the kernel would enlarge the trusted computing base behind eBPF's safety guarantee.
The verifier checks only the bytecode, so the kernel would have to trust the far larger compiler that produces the native code.
Such a compiler would also grow the kernel code that must be maintained and audited for each supported architecture.
A practical fix instead keeps the simple kernel JIT and targets the idiom share of the loss by letting the JIT emit a bounded set of hardware idioms directly.

Emitting the idioms from the simple kernel JIT must meet three constraints, which the next section addresses.
First, the verifier must still prove every program safe, which keeps eBPF's existing guarantee unchanged.
Second, adding an operation must not require rewriting the verifier or the JIT core, which keeps the kernel change small and one-time.
Third, one operation must map to a different native instruction on each architecture, which keeps verification of the operation architecture-independent.
Within these constraints the fix targets the operations C-RQ3 identifies.
The in-kernel execution cost (C-RQ1) and the bytecode-form residual (C-RQ2) stay out of scope, and \S\ref{sec:eval} measures how much of the JIT's loss the idioms carry.

%% file: tables/sec-3-micro-summary.tex
\begin{table}[t]
\centering
\small
\caption{Speedup over the kernel eBPF JIT on the 27 pure-bytecode microbenchmarks, as the geometric mean of median steady-state time. A path counts as faster on a benchmark when it beats the kernel eBPF JIT by more than 2\%.}
\label{tab:char-micro-summary}
\begin{tabular}{@{}llcc@{}}
\toprule
Platform & Path & Speedup & \# Cases Faster \\
\midrule
x86-64 & In-kernel native   & 1.55$\times$ & 24/27 \\
       & Userspace native   & 1.57$\times$ & 25/27 \\
       & Userspace LLVM-BPF & 1.53$\times$ & 25/27 \\
\midrule
ARM64  & In-kernel native   & 1.88$\times$ & 27/27 \\
       & Userspace native   & 1.98$\times$ & 27/27 \\
       & Userspace LLVM-BPF & 1.92$\times$ & 27/27 \\
\bottomrule
\end{tabular}
\end{table}

%% file: figures/sec-3-pure-bytecode-percase.tex
\begin{figure*}[t]
\centering
\includegraphics[width=\textwidth]{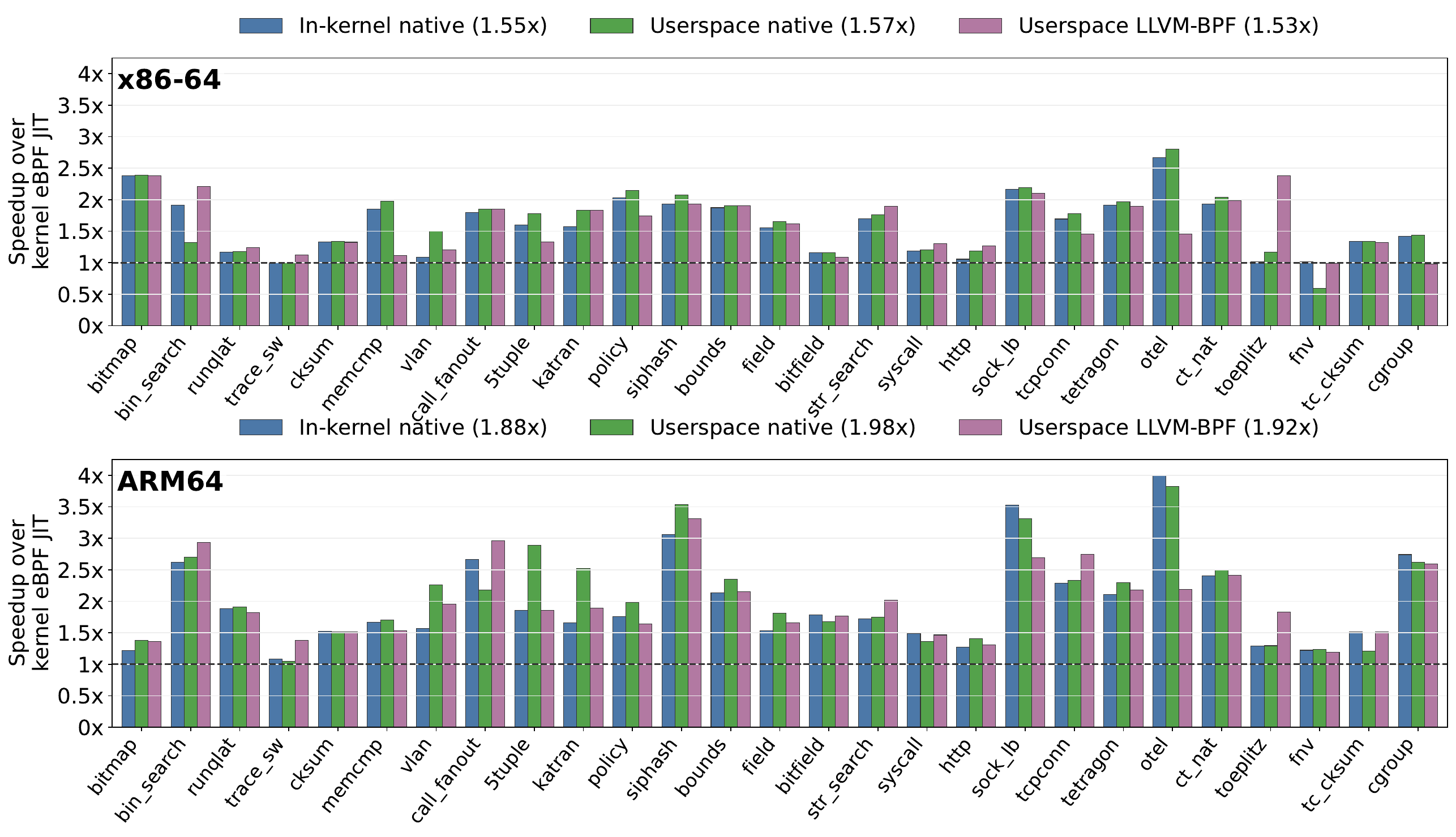}
\caption{Per-case speedup over the kernel eBPF JIT on the 27 pure-bytecode microbenchmarks, on x86-64 (top) and ARM64 (bottom). Each bar is one non-baseline path. The ARM64 paths come from separate runs, each over its own baseline.}
\label{fig:char-pure-percase}
\end{figure*}

%% file: figures/sec-3-jit-dump.tex
\begin{figure}[t]
  \centering
  \includegraphics[width=\linewidth]{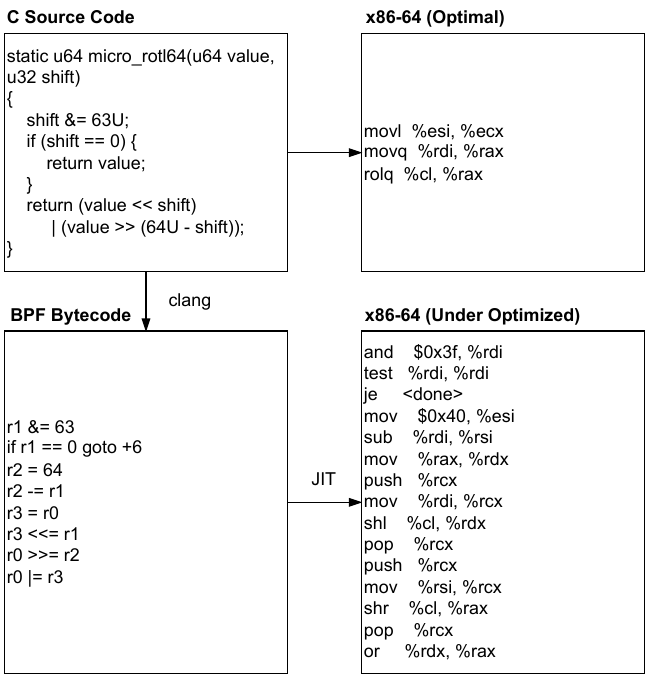}
  \caption{Compiling a 64-bit rotate with a variable shift. Direct compilation to x86-64 emits a single \texttt{ROL}. Through eBPF the rotate becomes eight bytecode instructions, which the per-opcode kernel JIT translates separately into 15 machine instructions (Linux 6.1, x86-64).}
  \label{fig:jit-dump}
\end{figure}

%% file: tables/sec-2-isa-gap.tex
\begin{table}[t]
\centering
\small
\caption{Hardware idioms and their instruction counts along each path, measured on x86-64 from real JIT dumps and compiler output. The second column names the native instruction on x86-64 / ARM64. Byte swap assumes a \texttt{MOVBE}-capable x86-64. The bulk-copy row is a 64-byte copy between non-overlapping regions.}
\label{tab:isa-gap}
\setlength{\tabcolsep}{3pt}
\resizebox{\columnwidth}{!}{%
\begin{tabular}{@{}llccc@{}}
\toprule
 & & \multicolumn{3}{c}{\# Instructions} \\
\cmidrule(lr){3-5}
Operation & Native instruction & Native & eBPF & Kernel JIT \\
\midrule
Rotate & \texttt{ROL}/\texttt{ROR} & 1 & 8 & 15 \\
Wide load & \texttt{MOV}/\texttt{LDR} & 1 & 22 & 22 \\
Conditional select & \texttt{CMP+CMOV}/\texttt{CMP+CSEL} & 2 & 3 & 4 \\
Byte swap & \texttt{MOVBE}/\texttt{REV} & 1 & 2 & 2 \\
Bulk copy (64\,B) & \texttt{MOVUPS}/\texttt{LDP+STP} & 4 & 16 & 16 \\
\bottomrule
\end{tabular}}
\end{table}

%% file: sections/4-bpfext.tex
\section{\tool: Extending the eBPF Compilation Pipeline}
\label{sec:mechanism}

\tool extends the eBPF compilation pipeline with new operations that the existing verifier checks and the kernel JIT emits directly.
Each operation has two forms: a \emph{proof sequence} of vanilla eBPF that the verifier checks and a \emph{native emit} that the CPU runs.
Figure~\ref{fig:pipeline} contrasts the pipeline with and without \tool.
A compile-time recognizer rewrites an operation's pattern into one or more \emph{extended instructions}, each a pair of a call and a sidecar that carries the operands.
Each call selects a \emph{descriptor} from a kernel module, and the descriptor supplies both forms for its instruction.
The same program therefore verifies as vanilla eBPF and runs as native code.
\tool meets the three constraints of \S\ref{subsec:char-challenges}: the safety guarantee stays intact up to one explicit obligation (\S\ref{subsec:safety}), the kernel core changes once (\S\ref{subsec:adding}), and verification stays architecture-independent (\S\ref{subsec:pipeline}).
\S\ref{sec:lang} then presents \lang, the hardware-idiom operations built with the mechanism.

\input{figures/sec-4-pipeline}

\subsection{The Compilation Pipeline}
\label{subsec:pipeline}

\noindent\textbf{Recognition.}
The recognizer matches the patterns of supported operations in userspace, before the program loads (\S\ref{subsec:llvm}).
For \lang's rotate, the pattern is the shift-and-or sequence of Figure~\ref{fig:jit-dump}.
The call and the sidecar reuse existing eBPF opcodes.
\tool therefore adds no new opcode.

\noindent\textbf{Verification.}
The verifier checks an extended instruction by lowering it to vanilla eBPF.
The proof sequence is a fragment of vanilla eBPF with the same semantics as the extended instruction.
A lowering stage substitutes the proof sequence for the extended instruction before the main analysis.
The verifier then analyzes the lowered program with its existing transfer functions.
After the analysis succeeds, a restore stage reinstates the extended instructions for the JIT.

\noindent\textbf{Execution.}
The JIT dispatches each extended instruction to the descriptor's native emit for the running architecture.
For \lang's rotate, the native emit is a single \texttt{ROL} on x86-64 or a single \texttt{ROR} on ARM64.
The exact proof sequence, like the native emit, can differ across architectures.
Every proof-sequence variant stays vanilla eBPF.
Verification therefore stays architecture-independent.
When a descriptor provides no native emit for the architecture, the extended instruction falls back to its proof sequence, which the JIT compiles as vanilla eBPF.
The interpreter has no dispatch for extended instructions.
A kernel without a JIT therefore rejects programs that carry extended instructions.

\subsection{Adding an Operation}
\label{subsec:adding}

A new operation needs only descriptors in a kernel module and a pattern in the recognizer.
A descriptor is a small kernel structure with two author-supplied parts, the proof-sequence routine and the per-architecture native emits.
A family of descriptors implements an operation, one descriptor per variant, such as width, condition, or architecture.
A kernel module registers its descriptors with the kernel core.
\tool's one-time change to the kernel core stays the same for every operation (\S\ref{sec:implementation}).

The mechanism leaves an operation's size open.
A proof sequence may span a few instructions or an entire region within the rules of \S\ref{subsec:safety}.

\subsection{Safety and the Trusted Base}
\label{subsec:safety}

\tool does not add operation-specific abstract semantics to the verifier.
Everything the recognizer and the module contribute reaches the verifier inside the lowered program.
The verifier re-checks the lowered program on every load, with the same analysis as for vanilla eBPF.
A faulty rewrite or proof sequence that violates safety therefore fails verification like any unsafe program.

The verifier's verdict on the lowered program carries over to the restored program.
The restore stage collapses each \emph{proof region}, the span its proof sequence occupies, back into its extended instruction and changes nothing else.
For the collapse to preserve the verdict, a region must stay single-entry and single-exit, free of nested extended instructions, calls, exits, and backward jumps.
Control then enters a region only at its top and leaves only past its end, like a single instruction.
The kernel validates proof-region well-formedness at load time, including single entry, single exit, no nested extended instructions, and no calls or exits inside a region.
The kernel also validates that no jump enters or leaves a region except at its boundary.

The trusted enforcement path therefore consists of the existing eBPF verifier and JIT plus a small, generic \tool patch that validates proof regions, lowers extended instructions before verification, restores them after verification, and dispatches their native emits (\S\ref{sec:implementation}).
The per-operation native emit is the one component whose correctness \tool cannot check at load time.
Every operation must therefore supply evidence that its native emit has the same eBPF-visible effect as its proof sequence.
For \lang, the Lean 4 proofs of \S\ref{sec:formal-verification} supply the evidence.
The userspace recognizer is outside the safety boundary: an incorrect rewrite produces a program the verifier re-checks and rejects or accepts on its own merits.

%% file: figures/sec-4-pipeline.tex
\begin{figure*}[t]
\centering
\includegraphics[width=0.9\textwidth]{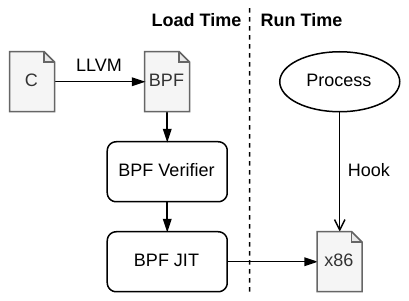}
\caption{The original eBPF pipeline (orange) and the \tool pipeline (green). The original pipeline verifies and JIT-compiles the bytecode unchanged, producing under-optimized native code. With \tool, the compile-time recognizer rewrites supported operations into extended instructions, and the \lang modules supply each operation's proof sequence to the verifier and its native emit to the JIT, producing optimized native code.}
\label{fig:pipeline}
\end{figure*}

%% file: sections/5-kinsn.tex
\section{\lang: Hardware-Idiom Operations}
\label{sec:lang}

With \tool, we build \lang, a set of seven operations for hardware idioms.
Each operation targets an idiom for which the eBPF instruction set has no single-instruction form.
The rotate, for example, covers its 64-bit and 32-bit variants through separate descriptors.
The section closes with the Lean 4 proofs that supply the evidence \S\ref{subsec:safety} requires.

The seven operations are a deliberate choice within the space that \S\ref{subsec:adding} leaves open.
The idioms of the kind \S\ref{subsec:char-idiom} characterizes are few.
A handful of modules therefore covers the kernel JIT’s clearest losses.
Each proof sequence stays a small fragment of the program.
The equivalence evidence each operation owes covers only that fragment.
The compiler emits six of the seven idioms as fixed bytecode patterns, which the recognizer matches in place.
The prefetch alone has no eBPF form of any length.
The recognizer instead inserts the prefetch at sites it selects, with a no-op proof sequence.

Table~\ref{tab:descriptors} lists the seven operations, each with its proof sequence and its native emit on x86-64 and ARM64.
The native emits differ across architectures, in shape and in availability.
A conditional select compiles to one extended instruction on x86-64, emitting the \texttt{CMP+CMOV} pair, and to two extended instructions on ARM64, one emitting \texttt{TST} and one emitting \texttt{CSEL}.
The byte swap, fused with its load, splits the same way on ARM64, into an \texttt{LDR} and a \texttt{REV}.
One operation, the address computation \texttt{LEA}, exists on x86-64 alone.

\input{tables/sec-3-descriptors}

\noindent\textbf{Formal verification.}\label{sec:formal-verification}
The dual lowering of a \lang: one path expands to vanilla eBPF for the verifier, the other emits native code in the JIT, imposes a proof obligation.
The native sequence that the CPU actually executes must produce the same eBPF-observable architectural state as the eBPF expansion.
We therefore prove, in Lean 4, that for each supported \lang operation, the eBPF expansion and the corresponding native sequence produce the same eBPF-observable architectural state.
For register-only operations such as rotate, conditional select, and bit extract, the refinement reduces to equality on the destination register.
For memory-touching operations such as byte swap (fused with its load) and bulk copy, the proof obligation is equality on the projected architectural state over live registers and the accessed memory, leaving all other eBPF-visible memory unchanged.
Prefetch is modeled as an architectural no-op; microarchitectural effects such as cache state are outside the semantic model.
We give each ISA a small-step register-transfer semantics; a program denotes the fold of its per-instruction transitions.
The native instruction’s semantics is ported from the existing formalization~\cite{lnsym}.
The cross-ISA proof is factored through an intermediate hardware-independent specification.
The refinement is stated for verifier-accepted proof sequences and verifier-safe initial states.
For each \lang operation, we define a specification over the relevant architectural state and discharge ISA-local correctness lemmas stating that the eBPF expansion and the native sequence each refine the specification.

%% file: tables/sec-3-descriptors.tex
\begin{table}[t]
\centering
\small
\caption{The seven \lang operations, each with the canonical proof sequence and the native emit per architecture. Exact proof sequences vary by descriptor and architecture (\S\ref{sec:mechanism}). A dash marks an architecture with no native emit, where the operation falls back to its proof sequence. Each native-emit cell lists the instructions emitted for one operation instance. Table~\ref{tab:isa-gap} counts the instructions along each compilation path for the idioms the two tables share, and the native compiler's instruction choice there can differ from the native emit here.}
\label{tab:descriptors}
\begin{tabular}{@{}llll@{}}
\toprule
 & & \multicolumn{2}{c}{Native emit} \\
\cmidrule(lr){3-4}
Operation & Proof sequence & x86-64 & ARM64 \\
\midrule
Rotate & two shifts + or & \texttt{ROL} & \texttt{ROR} \\
Conditional select & branch + move & \texttt{CMP+CMOV} & \texttt{TST+CSEL} \\
Bit extract & shift + and & \texttt{BEXTR} & \texttt{UBFX} \\
Byte swap & load + byte swap & \texttt{MOVBE} & \texttt{LDR+REV} \\
Prefetch & no-op (\texttt{JA +0}) & \texttt{PREFETCHT0} & \texttt{PRFM} \\
Bulk copy & loads + stores & \texttt{MOV} & \texttt{LDP+STP} \\
LEA & shift + adds & \texttt{LEA} & -- \\
\bottomrule
\end{tabular}
\end{table}

%% file: sections/6-implementation.tex
\section{Implementation}
\label{sec:implementation}

We implement \tool in three parts.
The kernel core is a one-time change to the verifier and the JIT (\S\ref{subsec:kernel}).
A compile-time recognizer rewrites idioms in userspace before load (\S\ref{subsec:llvm}).
Each operation ships as one or more loadable kernel modules (\S\ref{subsec:descriptors}).

\subsection{Kernel Core}
\label{subsec:kernel}

The kernel core adds 929 lines to the Linux kernel.
Table~\ref{tab:kernel-loc} breaks the total down by component.

\input{tables/sec-5-kernel-loc}

\noindent\textbf{Instruction encoding.}
Each extended instruction is a pair of instructions, a call and a sidecar, shown in Table~\ref{tab:encoding}.
Both reuse existing eBPF opcodes, and a reserved \texttt{src\_reg} value marks the pair for \tool.
The call identifies the descriptor to use.
The sidecar carries a packed 52-bit payload with the operands.
The low four bits of the payload must form a valid eBPF register number.
The recognizer therefore re-encodes a payload whose low four bits exceed the largest register number.
The kernel reverses the re-encoding when reading the sidecar.

\input{tables/sec-5-encoding}

\noindent\textbf{Verifier lowering and restore.}
The verifier gains a lowering stage and a restore stage around its existing analysis.
The lowering stage traverses the program backward and replaces each extended instruction with the proof sequence from its descriptor.
The stage records each original pair for restoration.
The verifier then runs unchanged on the lowered program.
The restore stage reinstates the extended instructions after verification succeeds.
With proof-sequence validation and the supporting bookkeeping, the changes to the verifier total 487 lines, the largest single component.

\noindent\textbf{JIT dispatch.}
The x86-64 and ARM64 JIT backends each gain a dispatch case for extended instructions.
On reaching an extended instruction, the JIT looks up the descriptor and decodes the sidecar payload.
The JIT then calls the descriptor's native emit for the running architecture.
When a descriptor has no native emit for the architecture, the kernel rewrites the site back to its proof sequence during load-time fixups, before the JIT runs.
The dispatch logic is 97 lines on x86-64 and 121 lines on ARM64.

\noindent\textbf{Module registration.}
A loaded module registers its descriptors for lookup by the verifier and the JIT.
Registration reuses two existing kernel facilities, BTF, the kernel's type metadata for eBPF objects, and kfuncs, kernel functions that eBPF programs may call.
Each module exports a stub kfunc as its BTF anchor.
The module lists its \texttt{struct bpf\_einsn} descriptors in a parallel array.
The kernel caches the descriptors per program.
The BTF and registration changes are 139 lines.
The new \texttt{struct bpf\_einsn} type and helpers add 85 lines of headers.

\subsection{Compile-Time Recognizer}
\label{subsec:llvm}

The recognizer rewrites a program's idioms into extended instructions before load, through two paths.
A program built without the recognizer, or one whose idioms the recognizer does not match, produces the same vanilla eBPF bytecode and runs through the unchanged pipeline.

\noindent\textbf{Selector.}
The first path is a selector in the LLVM eBPF backend.
The selector matches idioms during instruction selection and emits the call and sidecar in place of the decomposed bytecode.
Counted over the backend's \tool commits, the selector adds 2{,}126 lines, half in a new selection module.

\noindent\textbf{Bytecode recovery.}
The second path is a bytecode-recovery pass that re-matches idioms on already-compiled bytecode.
The pass covers bytecode the selector never sees or cannot match, such as programs compiled by a separate toolchain.
The implementation is a standalone userspace tool of 6{,}967 lines.
The pass checks that each rewrite preserves the program's semantics, for example that a temporary register the idiom overwrites is dead afterward.
Even a wrong rewrite leaves the program safe, because the verifier re-checks every lowered program (\S\ref{subsec:safety}).

\subsection{Operation Modules}
\label{subsec:descriptors}

Each operation is packaged as one or more kernel modules, separate from the kernel core.
A module supplies an operation's proof-sequence routine, its native emits, and the BTF registration that exposes both.

Each module covers one native instruction on one architecture.
An operation that targets a single architecture contributes modules on that architecture only, which is why the per-architecture counts differ.
The module tree holds 14 modules on x86-64 and 11 on ARM64, 9{,}765 lines in all, including shared helper headers, test-only modules, and operations beyond \lang's seven (Table~\ref{tab:descriptors}).

%% file: tables/sec-5-kernel-loc.tex
\begin{table}[t]
\centering
\small
\caption{The kernel core is a 929-line addition, measured as lines added on the \tool kernel branch over mainline Linux. The total excludes 11 lines of disassembler support and tools-side header sync.}
\label{tab:kernel-loc}
\begin{tabular}{@{}lc@{}}
\toprule
Component & Lines of code \\
\midrule
Verifier (lowering, restore, validation) & 487 \\
BTF and registration & 139 \\
JIT dispatch (ARM64) & 121 \\
JIT dispatch (x86-64) & 97 \\
Headers & 85 \\
\midrule
Total kernel core & 929 \\
\bottomrule
\end{tabular}
\end{table}

%% file: tables/sec-5-encoding.tex
\begin{table}[t]
\centering
\small
\caption{Extended-instruction encoding. A reserved \texttt{src\_reg} value marks the pair for \tool, a \texttt{BPF\_CALL} for the call and a \texttt{MOV64\_K} for the sidecar. The notation [h:l] gives a field's position within the 52-bit sidecar payload.}
\label{tab:encoding}
\begin{tabular}{@{}llcp{0.50\columnwidth}@{}}
\toprule
Insn & Field & Bits & Meaning \tabularnewline
\midrule
Call & \texttt{imm} & 32 & \raggedright BTF ID of the operation's stub kfunc \tabularnewline
 & \texttt{off} & 16 & \raggedright Index into \texttt{fd\_array} for the module's BTF file descriptor \tabularnewline
\midrule
Sidecar & \texttt{dst\_reg} & [3:0] & \raggedright Destination register or per-descriptor form tag \tabularnewline
 & \texttt{off} & [19:4] & \raggedright Offset or auxiliary field per descriptor \tabularnewline
 & \texttt{imm} & [51:20] & \raggedright Immediate value or configuration per descriptor \tabularnewline
\bottomrule
\end{tabular}
\end{table}

%% file: sections/7-evaluation.tex
\section{Evaluation}
\label{sec:eval}

This section evaluates \lang, the hardware-idiom instantiation of \tool.
\tool provides the extension mechanism, while \lang provides the concrete descriptor families used in our experiments.
\lang operates below the instruction-set boundary, complementary to bytecode optimizers~\cite{xu2021synthesizing,mao2024merlin,zhu2025epso} that cannot emit native idioms the ISA lacks.
We use two workload classes.
First, we reuse the 27 pure-bytecode microbenchmarks from \S\ref{sec:characterization} to isolate instruction-selection effects.
Second, we use production datapath applications to test whether those effects matter in real BPF loading and execution paths.
We answer four questions:

\begin{itemize}
\item \textbf{RQ1:} How effectively does \lang improve generated native-code efficiency across architectures?
\item \textbf{RQ2:} How does \lang perform on production applications?
\item \textbf{RQ3:} How sensitive is \lang to family-selection policy?
\item \textbf{RQ4:} How does \tool compare with a native-in-kernel upper bound?
\end{itemize}

\paragraph{Evaluation environment.}

We use two evaluation platforms.
The x86-64 experiments run in a KVM virtual machine with 8 vCPUs and 64\,GB RAM on an Intel Xeon Silver 4210R @ 2.40\,GHz host, using Linux 7.0.0-rc2+.
The ARM64 experiments run on an AWS t4g.small instance (2 vCPUs, 2\,GiB RAM, AWS Graviton2), using a \lang-enabled Linux 7.0.0-rc2+ build.
These platforms cover both the microbenchmark suite and the production-application experiments.
For matched application comparisons, the stock and \lang runs use the same kernel image and differ only in the loaded program form or \lang policy.

\paragraph{Measurement conventions.}
Unless otherwise stated, each experiment uses three measured runs and reports the median.
Execution-time and throughput ratios are normalized to the stock kernel eBPF JIT, so values above 1$\times$ indicate faster execution or higher throughput.
BPF-cost ratios report post/baseline cost, so values below 1$\times$ indicate lower BPF runtime cost.
For BPF-counter results, we report post/baseline ns/run for retained counter rows.
For each retained row, we compute \texttt{run\_time\_ns\_delta}/\texttt{run\_cnt\_delta}, drop rows with fewer than 100 runs in either phase, and pair rows by program name, program type, and occurrence index.

\subsection{RQ1: x86-64 and ARM64 Microbenchmark Signal}
\label{subsec:micro_benchmark}

\input{figures/sec-6-kinsn-micro-rq1}

\lang reduces generated native-code size on both evaluated architectures. Relative to the stock kernel eBPF JIT, generated kernel JIT code shrinks to 0.772$\times$ on x86-64 KVM and 0.879$\times$ on ARM64 AWS, corresponding to geomean reductions of 22.8\% and 12.1\%, respectively.

Figure~\ref{fig:eval-kinsn-micro} reports the corresponding execution-time results. On x86-64 KVM, the \lang-enabled run achieves a 1.242$\times$ geomean speedup over the stock kernel eBPF JIT, reducing execution time by 19.47\%. On ARM64 AWS, the \lang-enabled run applies all matched \lang sites, with 308/308 sites in the median sample and 924/924 site instances across all samples. Over the 27 \lang-bearing benchmarks, it achieves a 1.222$\times$ geomean speedup over the stock kernel eBPF JIT, reducing
execution time by 18.17\%. Both architectures produce zero correctness mismatches. On the microbenchmarks, the 1.242$\times$ speedup recovers 42\% of the characterization's 1.57$\times$ eBPF-native gap on x86-64 $((1.242{-}1)/(1.57{-}1))$.

The per-benchmark distributions suggest different sources of benefit across architectures. On x86-64, speedups are more closely associated with native code-size reduction, which is consistent with \lang replacing conservative stock-JIT instruction sequences and reducing dynamic instruction overhead. On ARM64, code-size reduction alone is less explanatory: the largest gains appear when the selector reaches target-specific idioms such as rotate/extract operations and packet-field load fusion. For example, \texttt{siphash\_rotate64\_\allowbreak mixer} reaches a 1.902$\times$ speedup on ARM64 with dominant extract-style sites, while networking-oriented microbenchmarks such as Cilium socket load balancing, BCC TCP tuple filtering, and Cilium CT/NAT tuple rewriting achieve speedups of 1.691$\times$, 1.624$\times$, and 1.482$\times$, respectively. These results suggest that \lang is not merely a code-size optimization. Rather, it acts as an architecture-sensitive specialization mechanism whose effectiveness depends on the stock JIT baseline, the selected descriptor family, and whether the selected sites lie on performance-critical execution paths.

The regressions also reveal the current boundary of the selector. \texttt{bpftrace\_comm\_key\_fnv\_hash} regresses to 0.862$\times$ on ARM64, while \texttt{bitmap\_popcount\_scan} regresses to 0.966$\times$ despite having applied \lang sites. These cases indicate that \lang is not uniformly beneficial and motivate descriptor-level gating or lightweight cost-model guidance, rather than blindly applying every matched \lang site.

\noindent\textbf{Load-time overhead.}
The \lang lowering and restore stages add two linear passes over the program during verification. Across all 62 microbenchmarks on x86-64 KVM, the kernel-side load time (\texttt{object\_load\_ns}, covering verification and JIT) shows a 0.99$\times$ geomean ratio relative to the stock kernel, i.e., no measurable overhead. For programs with \lang sites, the end-to-end compile time (including the userspace recognizer) increases by 1.4--2.4$\times$, but remains sub-millisecond in absolute terms.

\subsection{RQ2: Real Packet-Processing Applications}
\label{subsec:corpus_application}

Microbenchmarks isolate individual instruction patterns, but packet-processing applications combine those patterns with maps, helper calls, tail calls, and application-specific datapath structure. We therefore evaluate \lang on two real eBPF packet-processing applications: Cilium and Katran. Cilium represents a large tail-call-heavy Kubernetes datapath, while Katran represents a compact high-rate XDP load balancer with a hot directly attached program. We measure end-to-end datapath throughput with BPF runtime statistics disabled to avoid measurement overhead.

\textbf{Cilium on x86-64.}
On x86-64 KVM, the default full-\lang policy improves Cilium's end-to-end datapath throughput by 1.074$\times$ over the stock kernel eBPF JIT. This run applies 4086 \lang sites with zero load-time skips or report errors. The applied sites span several families: LEA contributes 2346 sites, conditional select 385, endian fusion 766, extract 2, and bulk memory 587. This confirms that \lang reaches real bytecode patterns in Cilium's production datapath.

\textbf{Katran on ARM64.}
On ARM64 AWS, the conservative \lang policy applies 21 sites and improves Katran's end-to-end workload throughput by 1.073$\times$ over the stock kernel eBPF JIT. Because Katran's datapath is dominated by a single hot XDP program, the throughput signal is clean: the program that receives \lang sites is the program on the critical path.

Together, these two case studies show that \lang improves real packet-processing workloads on both architectures. The results raise a follow-up question: once \lang reaches real application bytecode, should it enable every matched family, or should it select families by workload-specific profitability?

\input{figures/sec-6-kinsn-micro-rq3}

\subsection{RQ3: Family-Selection Policy Sensitivity}
\label{subsec:policy_sensitivity}

\lang creates a policy surface above the \tool kernel mechanism. The kernel mechanism exposes a bounded set of verified descriptor families, but enabling a family is a performance decision: each matched site trades the dynamic work saved by a shorter native idiom against the cost and placement of the replacement in a particular datapath. A simple policy would maximize coverage, i.e., enable every implemented family and apply as many matched sites as possible. Figure~\ref{fig:app_policy_probes} illustrates this policy effect by comparing two paired policy perturbations on Cilium and Katran.

For Cilium, we start from the full x86-64 \lang policy used in RQ2. As shown in Figure~\ref{fig:app_policy_probes}, this policy applies 4086 sites and improves workload throughput to 1.074$\times$, while the BPF-cost ratio remains near neutral at 1.009$\times$. We then disable only the \texttt{bulk\_memory} family while keeping the other \lang families enabled. This reduces the number of applied sites from 4086 to 3512, but increases workload throughput further to 1.114$\times$. At the same time, the BPF-cost ratio rises to 1.062$\times$, so this result should not be read as saying that \texttt{bulk\_memory} is universally harmful. Rather, it shows that an additional family can increase syntactic coverage without improving the measured datapath: the saved instructions may not outweigh the replacement overhead, or they may not lie on the part of the datapath that dominates throughput.

Katran shows the same policy lesson more directly. Figure~\ref{fig:app_policy_probes} shows that the conservative ARM64 policy used in RQ2 applies 21 sites and improves Katran to 1.073$\times$ workload throughput, while reducing BPF cost to 0.941$\times$. To test whether simply increasing coverage is beneficial, we then force a coverage-max policy that enables every implemented ARM64 family and applies 62 sites. The result is worse despite the larger number of applied sites: workload throughput falls to 0.995$\times$, and BPF cost rises to 1.006$\times$. Because Katran is dominated by a single hot XDP path, this result gives a particularly clear counterexample to coverage maximization: adding more matched sites does not translate into either higher throughput or lower BPF cost.

Taken together, the figure rejects raw site count as the right objective for \lang policy. The useful objective is profitability: a descriptor family should be enabled when its replacements save enough dynamic work on the target workload to justify their cost, rather than simply because they increase the number of matched sites.

\subsection{RQ4: Native-in-Kernel Upper Bound}
\label{subsec:native_upper_bound}

The three RQs above evaluate \lang as an instantiation of \tool. We now place \tool in a broader design space by using \tool itself to load whole-program native replacements. In this experiment, each Cilium BPF program is compiled from the same C source along two paths: the proof sequence is the normal eBPF bytecode that the verifier checks, while the native emit is the same source compiled with LLVM \texttt{-O2} directly to native code. The verifier still checks the eBPF proof sequence on every load, but the equivalence between the two compilations is trusted rather than proved, so the application's native code becomes part of the TCB. The 2.358$\times$ upper bound is not comparable to the characterization's 1.57$\times$ microbenchmark gap, which measures pure-bytecode kernels in isolation rather than Cilium's full datapath.

We use Cilium for this upper-bound study because it is the same kind of datapath workload evaluated above. The native-in-kernel experiment uses the same x86-64 KVM application setup as the Cilium \lang runs. It is not a claim about Cilium control-plane throughput: it measures steady-state datapath throughput after endpoint setup, with runtime reload and regeneration controllers disabled and the Cilium agent paused during measured workload samples.

Relaxing the trust boundary exposes a much higher performance ceiling. Cilium throughput improves by 2.358$\times$ over stock eBPF, and retained BPF cost drops from 488.7~ns/run to 262.3~ns/run, a 1.86$\times$ BPF-counter speedup.

The performance gain comes with a larger trusted surface. The Cilium native run logs 113 native replacements and 22 manifest no-match pass-throughs. The native sidecar data also records 89 Cilium manifest objects across 8 native files. These numbers are not a full TCB LOC audit, but they show the scale of the code surface that the native path must trust or validate.

This places stock eBPF, \tool instantiated as \lang, and native-in-kernel execution at three points in the same design space. Stock eBPF is the most conservative point: it keeps the existing verifier/JIT trust boundary but pays the cost of a small virtual instruction set. Native-in-kernel execution is the performance ceiling under an assumed-safe model, but it trusts whole native replacements and the loader path. \lang occupies the middle ground: its 1.074$\times$ Cilium throughput gain recovers 5.4\% of the 2.358$\times$ native upper bound's gap $((1.074{-}1)/(2.358{-}1))$, compared with 42\% on the microbenchmarks, because real datapaths spend most of their time in helpers, maps, and tail calls that \lang does not target. \lang does not reach the native upper bound, but it provides controllable speedups through a bounded, kernel-defined instruction surface rather than trusting whole native replacement objects.

%% file: figures/sec-6-kinsn-micro-rq1.tex
\begin{figure*}[t]
\centering

\begin{subfigure}[t]{\textwidth}
\centering
\includegraphics[width=\textwidth]{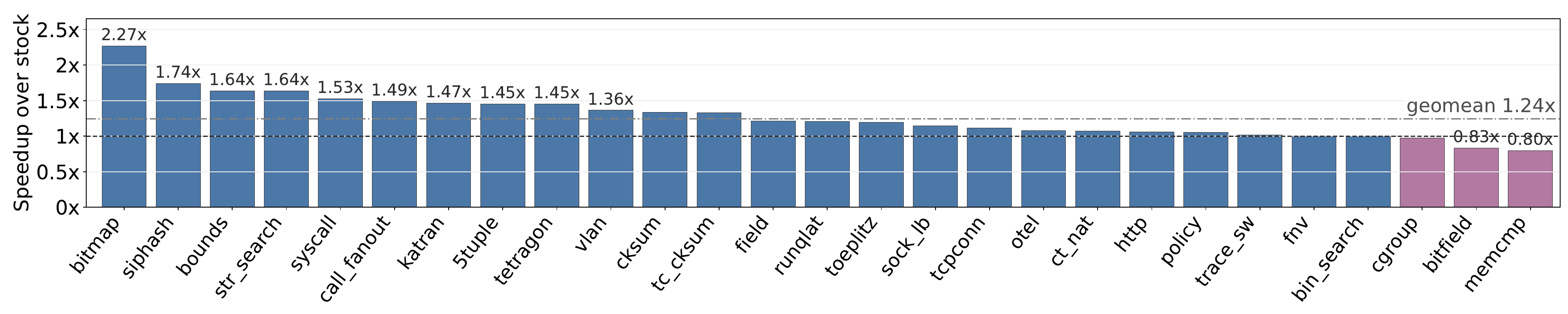}
\caption{x86-64 KVM \lang{}-enabled run.}
\label{fig:eval-kinsn-micro-x86}
\end{subfigure}

\vspace{0.6em}

\begin{subfigure}[t]{\textwidth}
\centering
\includegraphics[width=\textwidth]{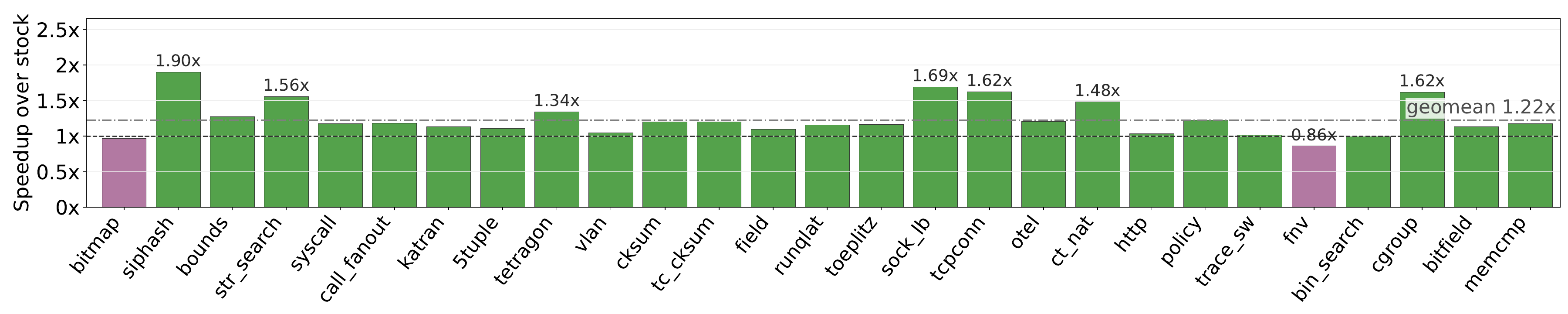}
\caption{ARM64 AWS \lang{}-enabled run.}
\label{fig:eval-kinsn-micro-arm64}
\end{subfigure}

\caption{Microbenchmark execution-time speedup over the stock kernel eBPF JIT.
Each benchmark is measured with three samples and
\texttt{INNER\_REPEAT=100000}; bars report speedup computed from median
\texttt{exec\_ns}, and higher is better. Panel~(a) reports the x86-64 KVM
\lang{}-enabled run. Panel~(b) reports the ARM64 AWS \lang{}-enabled run over
the 27 benchmarks with applied \lang{} sites. Red lines mark the geomean
speedup reported in each panel.}
\label{fig:eval-kinsn-micro}
\end{figure*}

%% file: figures/sec-6-kinsn-micro-rq3.tex
\begin{figure}[H]
  \centering
  \includegraphics[width=\columnwidth]{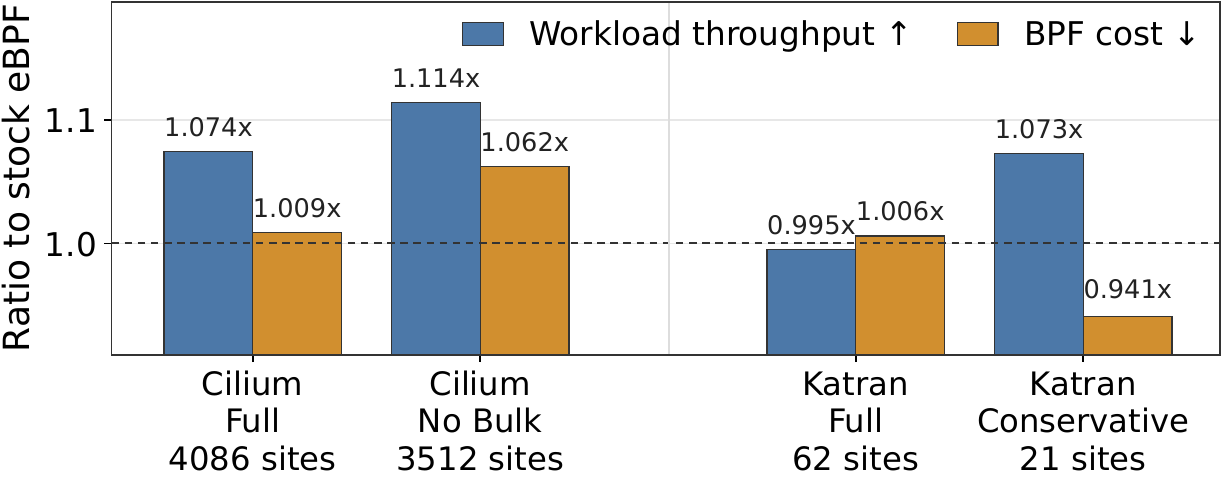}
  \caption{Application-level \lang{} policy probes. Workload throughput ratios
  are higher-is-better, while BPF cost ratios are lower-is-better. Cilium
  compares the full \lang{} policy with the no-bulk policy; Katran compares a
  more-sites policy with the selected-family policy. The dashed line marks the
  stock eBPF baseline.}
  \label{fig:app_policy_probes}
\end{figure}

%% file: sections/8-related-work.tex
\section{Related Work}
\label{sec:related}

\noindent\textbf{eBPF bytecode optimization.}
Bytecode optimizers improve eBPF programs within the existing instruction set.
K2~\cite{xu2021synthesizing} searches for shorter bytecode with stochastic synthesis.
Merlin~\cite{mao2024merlin} rewrites programs at the LLVM-IR and bytecode levels.
EPSO~\cite{zhu2025epso} applies rewrite rules from offline enumerative superoptimization.
ePass~\cite{xiang2025epass} lifts bytecode into SSA form for program transformation.
KFuse~\cite{kfuse} merges chains of eBPF programs into one program after verification.
\tool extends the set of operations available to a program.

\noindent\textbf{Extending eBPF's operations.}
The existing routes to new operations are kfuncs and instruction-set extensions.
A kfunc exposes a kernel function for eBPF programs to call~\cite{kfuncs}.
The verifier checks a kfunc call's arguments against BTF types, and the body runs as native code outside the verifier's analysis.
hXDP~\cite{brunella2020hxdp} extends the eBPF instruction set for FPGA targets, and the translated programs run outside the verifier's coverage.
The eBPF instruction set is now an IETF standard~\cite{rfc9669}, so a revision is a standardization effort.
A \tool operation shows the verifier its full semantics via a proof sequence in the lowered program (\S\ref{subsec:pipeline}).
A new operation ships as a kernel module plus a recognizer pattern, with no instruction-set revision (\S\ref{subsec:adding}).

\noindent\textbf{Relocating execution or the safety argument.}
Another line of work relocates eBPF execution or changes the foundation of its safety argument.
bpftime~\cite{zheng2025extending} runs eBPF programs in userspace.
eBPF for Windows~\cite{windows-ebpf} checks programs in userspace with the PREVAIL verifier~\cite{gershuni2019simple} and loads checked programs as signed driver modules.
Craun et al.~\cite{craun2023enabling} verify, compile, and sign programs on a separate host for embedded devices.
KFlex~\cite{dwivedi2024fast} moves only the extension's own memory safety and termination to runtime checks.
Rex~\cite{jia2025rex} drops the verifier and accepts extensions written in safe Rust.
MOAT~\cite{lu2024moat} and SafeBPF~\cite{soo2024safebpf} add runtime isolation beneath the verifier's static checks.
With \tool, programs stay in the kernel, under the existing verifier's load-time check.
Beyond \tool's generic enforcement core, the only per-operation addition to the trusted base is the native emit (\S\ref{subsec:safety}).

\noindent\textbf{Verified compilation and proof-carrying code.}
Verified compilation provides the techniques behind the \lang proofs.
CompCert~\cite{leroy2009formal} established mechanized semantic preservation for a realistic compiler.
Alive2~\cite{lopes2021alive2} checks refinement for each LLVM transformation.
Jitk~\cite{wang2014jitk} compiles classic BPF through a Coq-verified JIT that replaces the seccomp interpreter path.
Jitterbug~\cite{nelson2020specification} proves behavioral equivalence for per-instruction eBPF JIT translation with an SMT solver, and JitSynth~\cite{vangeffen2020jitsynth} synthesizes such verified JITs from interpreter semantics.
CertrBPF~\cite{yuan2022certrbpf} derives a verified C implementation of an eBPF interpreter from a Coq specification.
All of these systems target a fixed instruction set.
Each \lang operation instead includes its own Lean 4 proof that the native emit has the same eBPF-visible effect as its proof sequence (\S\ref{sec:formal-verification}).

Beyond the translators, machine-checked proofs also target eBPF's verification and loading steps.
Agni~\cite{vishwanathan2023verifying} verifies offline the eBPF verifier's range analysis, on whose soundness \tool's safety argument rests.
BCF~\cite{sun2025prove} supplies abstraction-refinement proofs from userspace, which the kernel checks at load time.
\lang's equivalence proofs stay offline, as development-time evidence that never enters the kernel.
Proof-carrying code (PCC)~\cite{necula1997pcc} is the closest ancestor of \tool.
Both admit untrusted code into the kernel only after an in-kernel check, but whereas PCC installs a new proof checker, \tool reuses the existing eBPF verifier.

%% file: sections/9-conclusion.tex



\section{Conclusion}
\label{sec:conclusion}

We presented \tool, a mechanism that extends the eBPF compilation pipeline with new operations the kernel JIT emits directly.
Each operation has a proof sequence that the existing verifier checks on every load and a native emit that the CPU runs.
Beyond the generic enforcement core, the native emit is the only per-operation addition to the trusted base.
With \tool, we built \lang, a set of seven hardware-idiom operations.
Lean 4 proofs show that each \lang native emit has the same eBPF-visible effect as its proof sequence.
\lang speeds up eBPF microbenchmarks by up to 24\% on x86-64 and 22\% on ARM64 over the unmodified pipeline, recovering up to 42\% of the gap to native code, and improves production application throughput by up to 12\%.
A new operation now ships as a kernel module, and the existing verifier analysis stays untouched.